\documentstyle[12pt]{article}
\begin{document}
\title{On quantization of Regge links}
\author{V. Khatsymovsky \\
 {\em Institute of Theoretical Physics} \\
 {\em Box 803} \\
 {\em S-751 08 Uppsala, Sweden\thanks{Permanent adress (after 15
November 1993): Budker Institute of Nuclear Physics, Novosibirsk
630090, Russia}} \\

 {\em E-mail address: khatsym@rhea.teorfys.uu.se\thanks{Permanent
E-mail address (after 15 November): khatsym@inp.nsk.su}}}
\date{\setlength{\unitlength}{\baselineskip}
\begin{picture}(0,0)(0,0)
\put(9,13){\makebox(0,0){UUITP-23/1993}}
\put(8,12){\makebox(0,0)}
\end{picture}
}
\maketitle
\begin{abstract}
In quantum Regge calculus areas of timelike triangles possess
discrete spectrum. This is because bivectors of these triangles are
variables canonically conjugate to orthogonal connection matrices
varying in the compact group. (The scale of quantum of this spectrum
is nothing but Plankian one). This is checked in simple exactly
solvable model - dimensionally reduced in some way Regge calculus.
\end{abstract}
\newpage

Regge calculus formulation of general relativity (suggested by Regge
in 1961 \cite{Regge}; see recent review \cite{Will-rev}) is most
suitable one for quantizing it, because of countability of the set of
field variables. The role of the latter is played by linklengths. At
the same time, contrary to usual lattice, Regge manifold is itself a
particular case of Riemannian manifold and it's discretization step
is itself field variable. On the other hand, any Riemannian manifold
can be approximated (in the sense of measure) with arbitrary accuracy
by Regge manifolds. Thus, not only Regge manifold is a particular
case of Riemannian one, but smooth Riemannian manifold is a limiting
case of Regge one. In this sense we may speak about two a'priori
equal in rights formulations; but if, for example, calculating
spectrum of the operator of length of Regge link we shall find it
separated from zero, this would mean that spacetime is discrete {\it
dynamically}, since a'priori nothing prevent from these observables
being arbitrarily close to zero thus reproducing continuum theory. It
is also clear that typical quantum of discrete spectrum (if any)
should be of the order of Plankian scale (the only dimensional
parameter at our disposal) and we can interpret this as if, while
having smooth manifold at macroscopic level, there were effectively
UV cut off at Plank mass in our theory.

In this note we consider simple model of Regge calculus where
quantization of areas of Regge triangles can be easily analysed. The
spectrum of these areas is found to be discrete for timelike
triangles. The idea of why it is so is rather simple: in the Palatini
form of continuum general relativity area tensors (bivectors) are
canonically conjugate to antisymmetric tetrad connections. In Regge
calculus connections become SO(3,1) matrices and those describing
rotations around timelike bivectors lay in the compact subgroup.
Corresponding conjugate areas should have discrete spectrum just as
angular momentum in usual quantum mechanics has.

In order to develop canonical quantization formalism one should make
one of coordinate, call it time $ t$, continuous by tending
dimensions of simplices in some direction to zero. To perform this
limiting procedure in a nonsingular way it is necessary to issue from
some regular arrangement of vertices w.r.t. $ t$-axis. Therefore we
consider the sequence of $ t=const$ 3D Regge manifolds, call these
leaves, separated by distances $ O(dt),~dt\rightarrow 0$ such that
for each vertex in any of the leaf there are both it's image and
pre-image at the distances $ O(dt)$ in the two neighboring leaves.
Besides that, we assume the same scheme of connection of different
vertices by links in the leaves. Then the full 4D Regge manifold is
supposed to result from these 3D leaves by triangulating the space
between each two neighboring ones by new, diagonal and infinitesimal
links. By infinitesimal and diagonal we call  simplices between the
two leaves if their measure (length, area, etc.) is $ O(dt)$ and
finite, respectively; those completely contained in the leaf will be
called leaf simplices. Infinitesimal links connect each vertex with
it's image in the neighboring leave. Diagonal link connects only
those vertices in the neighboring leaves whose images in one of these
leaves are connected by leaf link.

Note, we do not prescribe to the coordinate $ t$ definite sign of
invariant interval along it. Moreover, one should not do this since
in the full theory with a'priori equivalent coordinates such sign
should either be determined {\it dynamically} or be a free parameter.
%The word %'spacelike' is conventionally applied to the simplices
completely contained in one of 3D leaf; %by 'timelike' and 'diagonal'
we call simplices which contain vertices of the two leaves if their
%measure (length, area, etc.) is $ O(dt)$ or finite, respectively.

To describe the model, let us consider some points of continuous time
formalism in Regge calculus. In \cite{Kha} the canonical form of
Regge calculus in the connection representation has been found. The
coordinate $ t$ is continuous, and manifold is collection of 4-prisms
whose bases are 3D tetrahedrons filling each 3D section $ t=const$ of
such the manifold. There are local frames living in tetrahedrons,
connections $ \Omega$ (SO(3) matrices connecting these frames) and
antisymmetric area tensors (bivectors)

\begin{equation}\label{S-l-l}
S^{ab}={1 \over 2}(l^a_1l^b_2 - l^b_1l^a_2)
\end{equation}
living on triangles spanned by 4-vectors $ l^a_1,~l^a_2$. The
Lagrangian of the continuous time Regge calculus includes, among
others, the kinetic term,

\begin{equation}\label{kin}
L_{\dot{\Omega}}=\sum_{(ikl)}S_{(ikl)}*(\bar{\Omega}_{(ikl)}\dot{\Omeg
a}_{(ikl)}),~~~A*B\stackrel{\rm def}{=}{1 \over
4}\epsilon_{abcd}A^{ab}B^{cd}
\end{equation}
($ (ikl)$ is triangle with vertices $ i,~k,~l$), and the Gauss law
one,

\begin{equation}\label{Gauss}
L_h=\sum_{(iklm)}h_{(iklm)}*\sum_{{\rm cycle\, perm}\, iklm}

\varepsilon_{(ikl)m}\Omega^{\delta_{(ikl)m}}_{(ikl)}S_{(ikl)}
\Omega^{-\delta_{(ikl)m}}_{(ikl)},~~~
(\delta \stackrel{\rm def}{=}\frac{1+\varepsilon}{2}),\nonumber\\
\end{equation}
where $ (iklm)$ is the tetrahedron with vertices $ i,~k,~l,~m$, the $
h_{(iklm)}$ is antisymmetric tensor Lagrange multiplier, and $
\varepsilon_{(ikl)m}=\pm 1$ is sign function of vertices. Also, we
use sign convention for $ \epsilon_{0123}=+1$ and metric signature
(-,~+,~+,~+). There are also terms describing effect of induced
curvature in 3D section and those specifical for 4D Regge calculus
resulting from cancellations between contributions of closely located
leaf and diagonal triangles. (Each such triangle carries, generally
speaking, finite at $ dt\rightarrow 0$ curvature. Due to the
mentioned cancellation contribution of group of such triangles
between the two neighboring leaves into action is $ O(dt)$ thus
ensuring finite contribution to the Lagrangian at the same time
bringing new variables into theory.) However, these terms will not be
present in our simple model. Besides that, if bivectors of triangles
are considered as independent tensor variables, some bilinear
constraints should be imposed on them, e.g.

\begin{equation}\label{S-S}
S_{(ikl)}*S_{(ikl)}=0.
\end{equation}
This ensures the form (\ref{S-l-l}) for $ S$. There are constraints
of this type for $ S_1,~S_2$ of triangles sharing common edge and
also continuous time form of constraints expressing identity of
scalar product of two bivectors of the triangles having common edge
when calculated in different frames of the two neighboring
4-tetrahedrons sharing these triangles. These scalar product
constraints are assigned to ensure identity of lengths of those edges
of the 4-terahedrons which should coincide when building Regge
manifold of these 4-simplices. Again, the constraints not written out
will not be of interest for us.

Now let us describe the model itself. The simplifying suggestion is
to reduce dimensionality of each 3D leaf from 3 to 1 while keeping
rotational group SO(3,1) the same. Then we have the chain of vertices
$ i=...,1,~2,~3,...$ with SO(3,1) matrix $ \Omega^{ab}_i$ and
bivector $ S^{ab}_i$ defined at each vertex $ i$. Lagrangian includes
kinetic term and Gauss law:

\begin{equation}\label{mod-L}
L=\sum_{i}^{}{S_i*(\bar{\Omega}_i\dot{\Omega}_i)}+\sum_{i}^{}{h_i*(\Om
ega_iS_i\bar{\Omega}_i-S_{i+1})},~~~h^{ab}_i=-h^{ba}_i.
\end{equation}
Other terms describing effects of curvature are absent in two
dimensions. Besides that, we should impose on $ S_i$

\begin{equation}\label{mod-S-S}
S_i*S_i=0.
\end{equation}
Since the set of faces on which $ S_i$ are defined is disconnected
(vertices), all other bilinear constraints describing intersection of
different faces are absent.

Thus, our model is described by Lagrangian (\ref{mod-L}) and bilinear
constraint (\ref{mod-S-S}) on the chain of vertices. To avoid the
problem of boundary conditions, let us close the chain into circle,
i.e.

\begin{equation}
S_{i+N}\equiv S_i
\end{equation}
for some number $ N$. Denote $ S_1\equiv S$, find all other $ S_i$'s
by means of Gauss law and substitute back into $ L$. We find:

\begin{eqnarray}
L&=&\sum_{i=1}^{N}{S_i*(\bar{\Omega}_i\dot{\Omega}_i)}=S*(\bar{U}\dot{
U}),\\
&&U\stackrel{\rm def}{=}\Omega_N\Omega_{N-1}...\Omega_1,\nonumber
\end{eqnarray}
with the constraints (\ref{mod-S-S}) reduced to one that,

\begin{equation}\label{S-S-1}
S*S=0.
\end{equation}
Besides that, Gauss law results in

\begin{equation}
US\bar{U}=S.
\end{equation}
This means that $ U$ is the product of rotations around $ S$ and in
the plane of $ S$:

\begin{equation}\label{U-S}
U=\exp{\left( \alpha{^*\!S \over |S|}+\beta{S \over |S|}
\right)},~~~|S|^2\stackrel{\rm def}{=}-{1 \over
2}S_{ab}S^{ab},~~~^*\!S_{ab}\stackrel{\rm def}{=}{1 \over
2}\epsilon_{abcd}S^{cd}.
\end{equation}
Substituting this into $ L$ and disentangling exponents we have

\begin{equation}\label{L-S-a}
L=|S|\dot{\alpha}.
\end{equation}
On our two-dimensional manifold with SO(3,1) each $ |S|$ (this is
usual area up to $ \sqrt{-1}$) is real (literally, this looks like as
if we would set the space components of vectors $ l_1,~l_2$ in
(\ref{S-S-1}) {\it collinear} - this just corresponds to 3D Regge
submanifold degenerating into 1D structure). Now eq. (\ref{S-S-1})
does not constrain any dynamical variable and drops out. The $
\alpha$ is some rotational angle defined modulo $ 2\pi$, so we have

\begin{equation}\label{spec}
|S|=0,~1,~2~...
\end{equation}
for eigenvalues of $ |S|$.

Now two remarks are in order.

1. One could ask if it is correct to make classical transformations
leading from (\ref{mod-L}) to (\ref{L-S-a}) {\it before} applying
quantization procedure, since extra variables become operators in
quantum theory which can lead to new effects. In the unpublished
author's work \cite{Kha1} the full canonical analysis of the system
(\ref{mod-L}), (\ref{mod-S-S}) has been performed. At such the
analysis we rewrite the system in terms of the canonical variables $
\pi^{ab}_i,~\omega_{iab}$ where $ \omega_i$ is generator of $
\Omega_i=\exp{\omega_i}$ and $ \pi_i$ is some function of $
S_i,~\omega_i$ arising when writing kinetic terms in the form $
\pi^{ab}_i\dot{\omega}_{iab}$. In these variables 6-components (per
vertex) of Gauss law are not purely I class constraints, rather of
the 7-component system of Gauss law and (\ref{mod-S-S}) one can form
3 first class combinations; other four constraints are second class.
Adding three per vertex gauge conditions we can write out standard
form of path integral for the system subject to I and II class
constraints. Calculation of preexponential factor results in simple
power factor $ |S|^N$ for the path integral where integrations other
than $ d|S|d\alpha$ are already performed. This factor also can be
easily foreseen from dimensional considerations. This can be
interpreted as imaginary contribution to Lagrangian, $
-iN\ln{(|S|/C)}$, $ C$ being normalization constant. At such the
interpretation the only answer required by Hermiticity of action for
spectrum of $ |S|$ is that it consists of only one point, some
positive integer (to which $ C$ should also be equal):

\begin{equation}\label{}
|S|=n_0>0
\end{equation}

2. The role of closing the chain into circle might seem to be
decisive since namely this by means of Gauss law ensures special form
(\ref{U-S}) of dynamical variable $ U$. However, it is unlikely that
local effect of quantizing link was governed by global topology.
Indeed, for the gauge fixing mentioned in the first remark we can
take $ \omega^{0a}_i=0$ as admissible conditions \cite{Kha1}, so
variable conjugate to area still lays in the compact subgroup of
SO(3,1). Thus, the main role is played by Gauss law which generates
symmetries eliminating some degrees of freedom.

The word 'time' was used in this paper to denote continuous
coordinate playing the role of formal time in canonical quantization.
We, however, did not prescribe any sign to the invariant interval
along it. Moreover, in the theory of spacetime based on general
covariance one should not fix this sign by hand for it should be
determined {\it dynamically} or be a loose variable otherwise; only
the group of local rotations SO(3,1) is fixed. The word 'timelike' is
used, however, in it's usual sense just to denote link of negative
invariant interval or triangle with negative bivector squared. So, in
our model areas of timelike triangles are quantized with equidistant
spectrum proportional to Plankian scale (if gravity constant is
recovered). Moreover, this spectrum may consist of only one, nonzero
point. Of course, the question of what precisely happens in the full
four-dimensional theory can be answered only by solving the latter.

\bigskip
I am grateful to prof. A. Niemi, S. Yngve and personnel of Institute
of Theoretical Physics at Uppsala University for warm hospitality and
support during the work on this paper.

\end{document}